\documentclass{aa}
\usepackage{txfonts}
\usepackage{graphicx}

\begin{document}

\title{Interstellar Scintillation of PSR\ J0437$-$4715 on Two Scales
\thanks{Observations made with the US VLBA.  The National Radio Astronomy Observatory is
a facility of the National Science Foundation operated under cooperative agreement 
by Associated Universities, Inc.}}

\author{C.R. Gwinn\inst{1} \and C. Hirano\inst{1} \and S. Boldyrev\inst{2}}

\institute{Department of Physics,
              University of California, 
              Santa Barbara, CA 93106, USA
              \email{cgwinn@physics.ucsb.edu, hirano@physics.ucsb.edu}
           \and
              Department of Astronomy and Astrophysics,
              University of Chicago,
              Chicago, IL 60637, USA
               \email{boldyrev@flash.uchicago.edu}
               }

\date{Received 00.00.05 / Accepted 00.00.05 }

\abstract
{}
{We sought to determine the scale of scintillation in the interstellar plasma
of \object{PSR\ J0437$-$4715}.}
{We used the Very Long Baseline Array to obtain scintillation amplitude and phase data,
from dynamic spectra at 327 MHz.}
{We observe two scales of scintillation of pulsar PSR\ J0437$-$4715,
differing by more than an order of magnitude in scintillation bandwidth.
The wider-bandwidth scale of scintillation that we observe 
indicates less scattering for this pulsar than for other nearby pulsars,
other than \object{PSR\ B0950+08}.
}
{}


\maketitle

\section{Introduction}

In this paper we report measurement of the interstellar scattering by plasma along the line of sight to 
pulsar J0437$-$4715. This is one
of the closest pulsars.  It has distance $R = 150$\ pc, and its transverse velocity is
100 km/sec (van Straten \cite{Van01}).  
It is quite strong over a wide range of observing frequencies.
Its proximity, and its intensity, make it an important probe of the local interstellar plasma.

We describe observations of the scattering parameters of J0437$-$4715, made at 327\ MHz.
We find two scales of scintillation,
corresponding to two scales of structure in the pulsar's dynamic spectrum.
Multiple scales are not uncommon for nearby pulsars 
(see, for example, Stinebring et al. \cite{Sti01}, Hill et al. \cite{hil03}),
although scales separated by more than an order of magnitude
are unusual.
Comparison with other pulsars of similar distance and dispersion measure
shows that this pulsar has relatively less scattering than expected
on the basis of a model for homogeneous turbulence.
In this paper, we discuss the observations and comparison with other pulsars.  
We discuss comparison with previous observations of scintillation of J0437$-$4715, and 
interpretation of the measurements,
in an accompanying paper (Smirnova et al. \cite{smi06}), hereafter Paper 2).

\section{Observations and Analysis}

\subsection{Data Collection and Calibration}

We observed PSR\ J0437$-$4715 at two epochs using the Very Long Baseline Array
(VLBA) operated by the National Radio Astronomy Observatory (NRAO).
On 5 November 1996, we observed the pulsar over four 33-minute scans,
using 25-meter telescopes at Fort Davis, Mauna Kea, Pie Town, and
St. Croix, with a 32-MHz bandwidth centered at 332~MHz, and with a
250-kHz spectral resolution.  Between the scans, we observed
extragalactic sources for calibration purposes.  On 8 April 1999, we
observed the target for two 44-minute scans and one 33-minute scan,
using the telescopes at Fort Davis, Kitt Peak, Los Alamos, Mauna Kea,
Owens Valley, Pie Town, and St.~Croix, with a bandwidth of 32~MHz
centered at 328~MHz, and with a spectral resolution of 125~kHz. Mutual
visibility is short, particularly on long baselines, because the source lies
at low declination; this limited the useful time span of the observations. To
increase the signal-to-noise ratio, the data for this observation were
correlated using a gate which covered 12\% of the pulse period and
which isolated the central peak of the pulse profile.  Scans of a
calibrator source preceded and succeeded the observations of the
pulsar.  For both observations, we aggregated the 32-MHz bandwidth
using four 8-MHz intermediate-frequency (IF) bands.

We performed the first half of the calibration using NRAO's Astronomical Image
Processing System (AIPS).  The data from both observations were marred
by radio-frequency interference (RFI).  The second observation in
particular suffered badly from RFI, forcing us to eliminate the Kitt
Peak and Owens Valley antennas from the outset.  The interference also
tainted some measurements of system temperature.  The
system temperature should not change abruptly, so if an isolated
reading clearly departed from an otherwise smooth trend, we 
removed that value.  In two IFs, however, rampant corruption due to
RFI obliterated evidence of a smooth evolution in system temperature.
For these two IFs, we copied the system temperatures from an adjacent
IF.

We examined the remaining visibilities and flagged channels and time
ranges to eliminate data corrupted by very strong, narrow-band signals
typical of man-made sources.  We applied the usual corrections for the
effects of antenna parallactic angles, digital sampler bias, system
temperature, and gain calibration.  We then performed a manual phase
calibration by running a fringe fit on a thirty-second interval of
calibrator data to determine the electronic delay differences between
IFs, and we removed the resulting phase differences from the pulsar
and calibrator data.  The final procedure done using AIPS was a fringe
fit on each source to remove phase slopes due to the residual fringe
rate and delay.

To perform a bandpass calibration, we averaged, for each baseline, the
complex visibilities from a ten-minute interval of calibrator data.
Deviations from a flat amplitude and phase spectrum reflected the
non-ideal frequency response of the baseband converters.  The gain of
the filters drops off at the ends of each IF, significantly reducing
the signal-to-noise ratio in the first and last channel of each IF;
consequently, we omitted these channels in our subsequent analysis.
We then smoothed the data using boxcar averaging with a one-minute
window to further increase the signal-to-noise ratio.
We then exported the data from AIPS and continued calibration using custom software.

At our observing frequency, propagation through the ionosphere
significantly affects the visibility phases in both time and 
frequency (Thompson et al. \cite{Tho86}).  
Data from both observations exhibited broadband variations
in phase of order half a radian over a time scale of about fifteen
minutes, as would be expected for ionospheric path changes.
We approximated and removed these temporal variations using an eighth-order
polynomial fit to amplitude-weighted phase averages calculated for
each time point.

The ionosphere also introduces a nonlinear 
variation of phase with frequency (Thompson et al. \cite{Tho86}).  The fringe
fit removes only a linear dependence of phase on frequency.
Usually, the bandwidth is small enough compared to the observing
frequency that the higher-order terms are not important.  However, in
our case, the bandwidth was approximately 10\% of the observing
frequency, and the resulting 1\% second-order
correction would be of order a radian
for typical values of total electron content for the ionosphere.  We
fit the amplitude-weighted phase averages calculated for each channel
to a quadratic polynomial and used this polynomial to calculate the
appropriate phase to subtract from the visibilities in each channel.
This yielded the amplitude as a function of frequency, as shown for example
in Figure\ \ref{data} for the short Fort Davis-Pie Town baseline.

To verify that our calibration procedure did not introduce any
unintended biases, we calculated two-dimensional autocorrelation
functions (ACF) for each data set using the visibilities from a
calibrator source. As expected, we obtained results flat in both time
and frequency lags, except for a spike at small time and frequency
lags due to the effects of noise, broadened from a delta-function
by smoothing.

\subsection{Observational Results}

\subsubsection{Dynamic Spectra}

Initially, we had hoped to measure the angular broadening of the
pulsar by measuring the phase variations of the diffraction pattern;
however, several complications derailed our efforts.  Most
significantly, the scintillation bandwidth was much larger than
expected.  As the phase variations of the diffraction pattern have
this characteristic bandwidth (Desai et al. \cite{Des92}), they could not be
distinguished from ionospheric phase, phase slope, and curvature.
Moreover, the angular scale inferred from the scintillation bandwidth
is so small that we would not expect to detect phase variations on
any Earth-based baseline, at 327\ MHz or higher observing frequency.

We chose to
focus on the Fort\ Davis-Pie\ Town baseline, which had the least interference
and best signal-to-noise ratio.  Figure\ \ref{data} shows
plots of the visibility amplitudes for the pulsar on this baseline for
both epochs.  The white vertical lines correspond to the channels
omitted at the ends of each IF, and the white horizontal lines
correspond to breaks in the scan or to time ranges flagged because of
RFI.  At both epochs, we detected scintillation maxima that typically
persist for ten to twenty minutes and that span several IFs,
suggesting a characteristic width in frequency on the order of ten
megahertz.

\subsubsection{Scintillation Bandwidth and Timescale}\label{scint_bw_tscale}

\begin{figure}
\resizebox{\hsize}{!}{\includegraphics{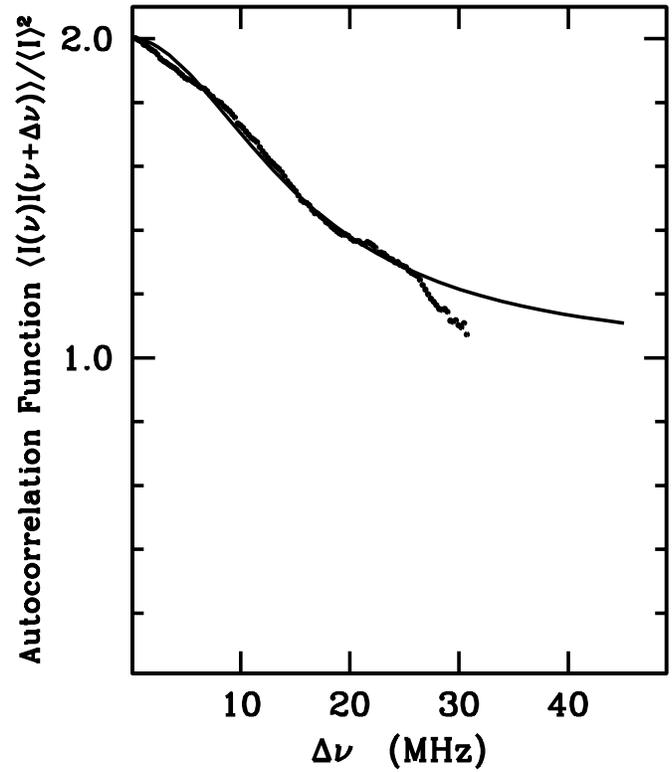}}
\caption{The composite ACF obtained by suitably combining the ACFs
from the two observations.  The best-fitting Lorentzian has a
characteristic width of $\Delta\nu_{ISS}=15.7\pm 0.2~\rm MHz$.}
\label{freq}
\end{figure}

We formed autocorrelation functions of the data shown in Figure\ \ref{data}.
Figure\ \ref{freq} shows an example, a composite autocorrelation function formed
for all the data.
We then fit models to the autocorrelation functions, using the form expected for Gaussian spectra:
a Lorentzian for differences in frequency and a Gaussian for differences in time
(see Gwinn et al. \cite{Gwi98}).
We found values for the decorrelation bandwidth of
$\Delta\nu =14.7\pm 0.5\ \rm{MHz}$ for the first epoch and
$\Delta\nu =16.5\pm 0.5\ \rm{MHz}$ for the second,
measured as the half-width at half maximum of the autocorrelation
function in frequency.  
A fit to a composite autocorrelation function,
using all of our data, yielded $\Delta\nu =15.7\pm 0.2\ \rm{MHz}$.
This composite function is the sum of the autocorrelation functions for the two epochs,
weighted by the number of samples.
The quoted errors are standard errors for the fits; 
we discuss the expected uncertainties further below.
The scintillation timescale $t_{ISS}$ was about 1000\ sec, as discussed below.
Table\ \ref{results_table} summarizes the results.

As Figure\ \ref{data} shows,
the number of independent samples with 
dimensions $\Delta\nu \times t_{ISS}$
is small, so that the accuracy of our measurement is
limited by number of samples. 
Phillips \& Clegg (\cite{Phi92}) argue that the fractional accuracy of such a measurement
is given by the number of independent samples, so that
\begin{equation}
{\sigma(\Delta\nu )}^2/{\Delta\nu} 
= 1/(N_t \times N_f )
= {{t_{ISS}}/T} \times {{\Delta\nu}/{B}} .
\label{phicalc}
\end{equation}
For our observations, $N_t=T/t_{ISS}\approx 14$, and 
$N_f=B/{\Delta\nu} \approx 2$.
We thus expect 
$\sigma(\Delta\nu )\approx (1/\sqrt{28})\Delta\nu \approx 3$\ MHz.

Because the total bandwidth is only twice the decorrelation bandwidth,
we used Monte Carlo simulations with a range of parameters to 
estimate errors as well.
Using the stationary-phase technique 
(Gwinn et al. \cite{Gwi98}),
we formed 1500 dynamic spectra with the same $N_t$ and $N_f$
as our observations, with ``true" decorrelation bandwidth evenly distributed
between 0 and 50\ MHz.
We then found the autocorrelation function for these synthetic spectra, 
and fitted for the decorrelation bandwidth,
as described above for the actual data.
We then found the distribution of ``true'' decorrelation bandwidth
that contributed to Monte Carlo results close to our measured value.
Thus, we inferred a parent distribution for our observed value.

We found that the parent distribution was sharply peaked near the measured
value of 16\ MHz, but had an extended tail toward larger values.
Of those spectra that yielded values near our measured value, 67\% came from the
interval between 13\ MHz and 23\ MHz.
The endpoints of this interval have about equal probability, according to our
Monte Carlo simulations; their asymmetry about the most probable 
value of 16\ MHz reflects the tail toward larger $\Delta\nu$.
The fact that we obtain values close to 16\ MHz independently,
from both of our observing epochs, 
provides additional evidence that 16\ MHz is the ``true" underlying value; we did not 
demand that our Monte Carlo simulations reproduce this additional fact,
to keep them manageably simple.
This simulation suggests that the uncertainty suggested by
Eq.\ \ref{phicalc} is approximately correct, with additional probability on the
side toward larger $\Delta\nu$.
The last line in Table\ \ref{results_table} indicates our adopted values for uncertainties,
from the Monte Carlo simulations, of $16^{+8}_{-3}$\ MHz. 

The Monte Carlo simulations indicate that the actual decorrelation bandwidth is 
unlikely to be about 1\ MHz, as would be expected from some previous measurements
(Paper 2).
To test this hypothesis, we formed 1000 spectra with true decorrelation bandwidth of 1\ MHz,
and then found the autocorrelation function and fitted value for the decorrelation
bandwidth. None approached 16\ MHz.  The largest value obtained was 4.7\ MHz.
We conclude that the probability of the decorrelation bandwidth actually being close to 1\ MHz
is much less than 0.1\%.

Similarly, we measured the scintillation timescale as the
half-width at half-maximum of the time correlation function with time.
The measured scintillation time scales were
$t_{ISS}=885\pm 9\ \rm{s}$ for the first epoch
and
$t_{ISS}=1630\pm 80\ \rm{s}$ for the second. 
The composite autocorrelation function yielded a time scale of 
$t_{ISS}=1020\pm 70\ \rm{s}$.
The expression of Phillips \& Clegg (\cite{Phi92}) 
would indicate that the true value was $1000\pm 200\ \rm{s}$.
Monte Carlo simulations of the parent distribution indicate
a confidence interval of $t_{ISS}=1000^{+350}_{-250}\ \rm{s}$.
The larger estimated errors 
we find from Monte Carlo simulations, for both $t_{ISS}$ and $\Delta\nu$,
may result from departures of the correlation function
from its expected form, for small $N_t$ and $N_f$.
We adopt the central value from the composite autocorrelation function in time, and the uncertainty
from Monte Carlo simulations or the expression of Phillips \& Clegg, as shown by the last line in
Table\ \ref{results_table}.

\subsubsection{Fine-Scale Scintillation}\label{fine_scale_sect}

\begin{figure}
\resizebox{\hsize}{!}{\includegraphics{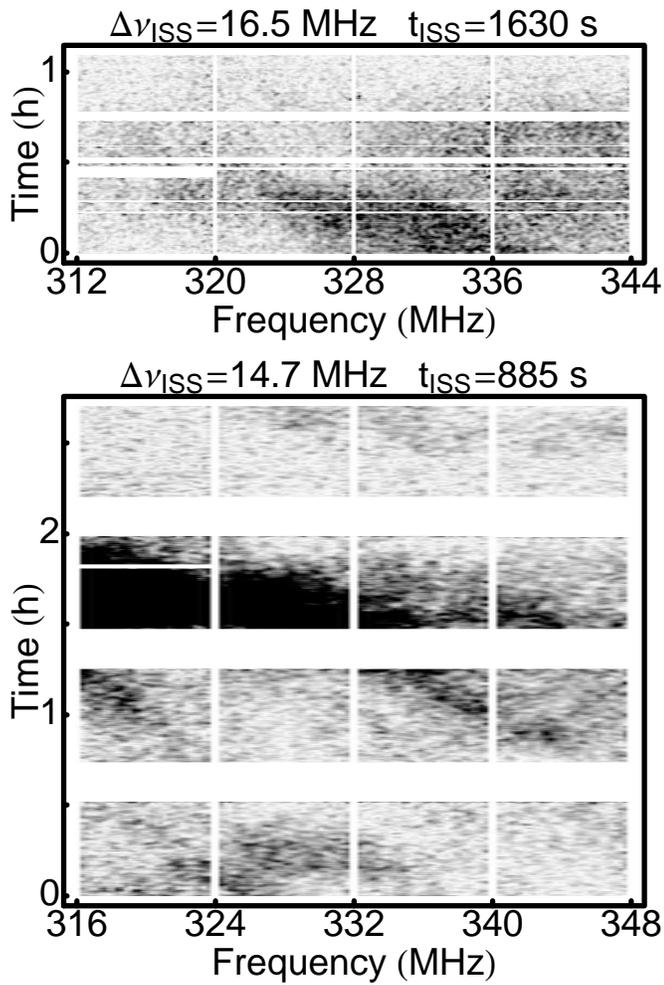}}
\caption{Visibility amplitude as a function of
frequency and time for observations of PSR\ J0437$-$4715 on the
Fort~Davis-Pie~Town baseline on 5 November 1996 (bottom) and 8
April 1999 (top).}
\label{data}
\end{figure}

\begin{figure}
\resizebox{\hsize}{!}{\includegraphics{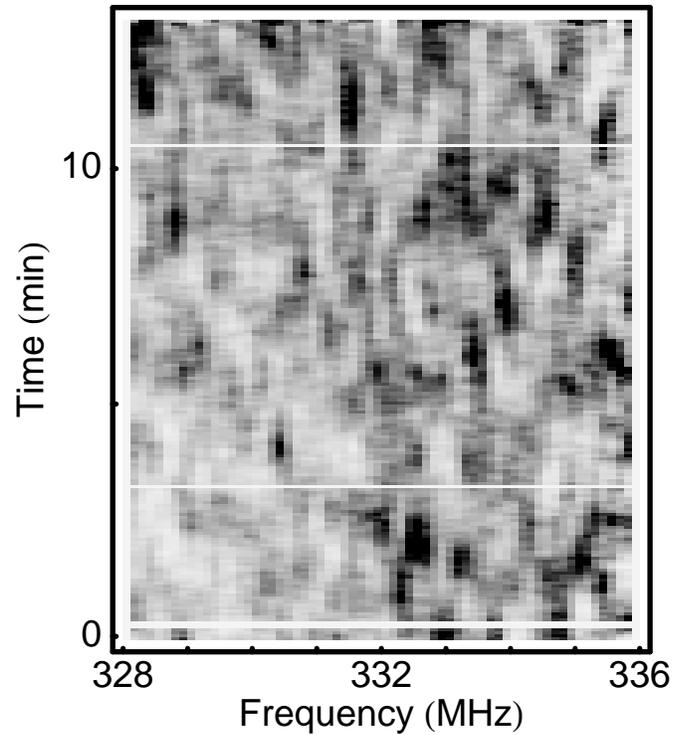}}
\caption{Variations in the visibility amplitude within a scintillation
maximum appear to occur over a frequency scale on the order of 1~MHz.}
\label{visib2}
\end{figure}

A finer-scale variation within the scintillation maxima is apparent
within the large-scale structures in Figure\ \ref{data}.  This variation
is not noise: it is a real modulation of intensity.
Figure\ \ref{visib2} shows an expanded view of part of a
scintillation maximum in Figure\ \ref{data}.
These finer-scale scintillations are also
visible in the autocorrelation function,
when the domain of integration is limited to 
a single peak of the scintillation spectrum in Figure\ \ref{data}.
Analysis indicates that they
have bandwidth $\Delta\nu\approx 0.5\  \rm{MHz}$
and time scales of 1 to 3 minutes.
The modulation index of this finer-scale structure
is $m_f\approx 0.2$.
These values are in approximate agreement with the scintillation parameters other groups
have reported for PSR\ J0437$-$4715,
as we discuss further in Paper 2. 

For the fine-scale scintillations, 
the autocorrelation function shows differences from the
forms described in \S\ref{scint_bw_tscale} above,
even when the reduced modulation index is taken into account.
These might arise from the modulation introduced by the larger-scale scintillation;
or by variations in the characteristic scales of the fine-scale scintillation, perhaps
similar to those described by by Gothoskar \& Gupta (\cite{got00}), but within a single observation.
The number of independent samples of the fine-scale scintillation is large enough
to have negligible contribution to the expected errors.
We adopt the standard errors for our fits,
as our best estimates.
Variations of the properties of small-scale scintillation 
with epoch, as Gothoskar \& Gupta observe,
could be greater than these errors, but we cannot
assess them from our data.
In practice, longer and greater-bandwidth observations of this pulsar,
or observations at lower frequencies where scales are smaller, are probably required to
understand statistics of the finer-scale scintillation.
We discuss a possible explanation of the fine- and wide-scale scintillation in Paper 2.

\section{Discussion}
\subsection{Scintillation Velocity}

The combination of scintillation bandwidth and time scale yields an estimated
velocity for the scintillation pattern, usually dominated by the velocity of the pulsar.
The relation
(Gupta et al. \cite{Gup94}):
\begin{equation}
V_{ISS} =  A_V \frac{\sqrt{\Delta\nu_{ISS}\, D\, X}}{f \, t_{ISS}}
\label{viss}
\end{equation}
connects scintillation velocity $V_{ISS}$ to the observables $\Delta\nu_{ISS}$ and $t_{ISS}$,
along with the distance of the pulsar $D$, a constant $A_V$,
observing frequency $f$,
and  $X$, the ratio of the distance from observer to scattering material to that
from scattering material to pulsar. For material halfway $X=1$;
for material close to the observer $X<1$. 

The 
calculated scintillation velocity $V_{ISS}$ varies 
along with the scintillation bandwidth.
Gothoskar et al. (\cite{got00}) and Nicastro et al. (\cite{nic95})
both report variation of the inferred $V_{ISS}$
by about a factor of 2,
as inferred from individual observations.
The $V_{ISS}$ obtained for our adopted values for the wide-band scintillation in Table\ \ref{results_table}, with 
$A_V=3.85\times 10^4\ {\rm{km\ s}^{-1}} \times ({\rm 1\cdot GHz\ 1\ sec})(\rm{1\ MHz\cdot 1 kpc})^{-1/2}$ (Gupta et al. \cite{Gup94}),
and an assumed distance of 150~pc (van Straten \cite{Van01}),
and $X=1$, is ${180^{+75}_{-50}}\ {\rm{km\ s}^{-1}}$. 
This is close to the average value of
$170\ {\rm{km\ s}^{-1}}$ reported by Johnston et al. (\cite{joh98}), 
and consistent with the $231\ {\rm{km\ s}^{-1}}$ reported by Gothoskar \& Gupta (\cite{got00}).
The proper-motion velocity is $100\ {\rm{km\ s}^{-1}}$ (van Straten et al. \cite{Van01}).
The values calculated from our two epochs separately, $200\ {\rm{km\ s}^{-1}}$ and $110\ {\rm{km\ s}^{-1}}$,
differ with marginal significance from the adopted value;
more extensive observations could reveal whether the time-variability 
observed by Gothoskar \& Gupta holds for the wide-band scintillation.
Gothoskar \& Gupta (\cite{got00}) inferred from the relatively high scintillation
velocity that the fractional distance of the scattering material,
$x$ (see Eq.\ \ref{viss}),
was larger than 1,
and surmised that the scattering material likely lies at the surface
of the Local Bubble.
As discussed in Paper 2,
some evidence suggests that the scattering lies in a thin layer,
$\approx 10$\ pc from the Sun.

\subsection{Comparison with Other Pulsars}

\begin{figure}
\resizebox{\hsize}{!}{\includegraphics{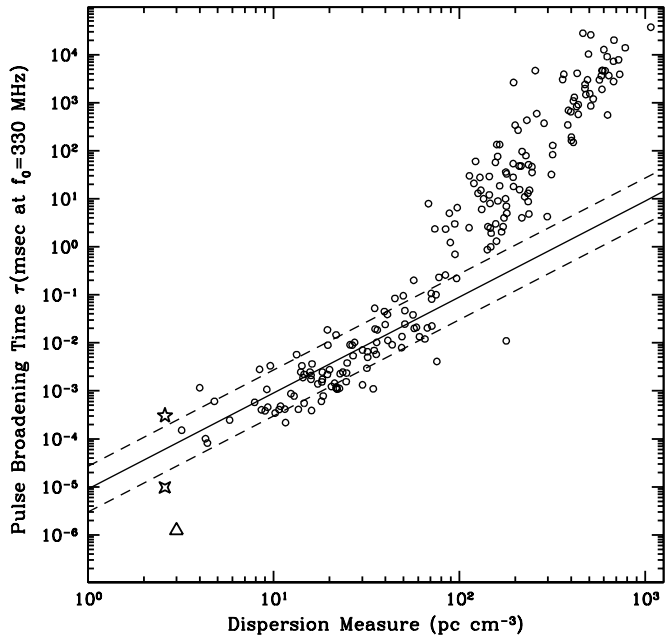}}
\caption{Pulsar pulse broadening time plotted with dispersion measure, 
adapted from
Pynzar' \& Shishov (\cite{pyn97}).
As they discuss, the solid line shows the scaling of pulse broadening with distance
expected for a uniform medium.
Stars show the wide-band (4-pointed) and fine-scale (5-pointed)
values we measured for PSR\ J0437$-$4715;
error bars and variation between epochs lie within the symbols.
Triangle shows value for PSR\ B0950$+$08 measured by Phillips \& Clegg (\cite{Phi92}).
Data for other pulsars from Cordes \& Lazio (\cite{cor02}).
}
\label{elbow}
\end{figure}

PSR\ J0437$-$4715 has among the lowest dispersion measures of strong pulsars,
and it is perhaps unsurprising that the scintillation 
bandwidth is among the broadest observed, scaled to a single frequency.
Comparisons with other pulsars are usually done by the typical pulse
broadening time $\tau$ (actually the time required for a broadened pulse to reqch $1/e$ of its maximum) (Pynzar' \& Shishov \cite{pyn97}).
The scintillation bandwidth is related to the typical pulse broadening time,
$\tau$, by the equation:
\begin{equation}
\Delta\nu_{ISS} \tau = C_1/(2\pi)
\end{equation}
where $C_1$ is a constant of order unity
(see Taylor et al. \cite{tay93}, Lambert \& Rickett \cite{lam99}).
We follow Pynzar' \& Shishov (\cite{pyn97}) and adopt $C_1=1.0$, and 
scale to observing frequency $f_0=330$\ MHz using $\tau\propto f^{-4.4}$.
Figure\ \ref{elbow} compares the scintillation parameters of 
PSR\ J0437$-$4715 with those of other pulsars.
The broader-$\Delta\nu_{ISS}$ scattering of PSR\ J0437$-$4715 falls below the scaling found by  
Pynzar' \& Shishov (\cite{pyn97})
for nearby pulsars, as does the scattering of PSR\ B0950$+$08 (Phillips \& Clegg \cite{Phi92}). 
Interestingly, 
some earlier measurements of the scintillation properties of 
PSR\ B0950$+$08 
found spectral structures that would indicate much larger $\tau$, 
placing it near the line extrapolated from other nearby pulsars.
Those measurements also indicated 
frequency and time 
structures unexpected for scintillation, and large variations with observing epoch
(Roberts \& Ables \cite{rob82}).
The fine-scale structure we observe for PSR\ J0437$-$4175 lies above this extrapolated line,
although observations of this object by others yield values close to the line.
We discuss observations of PSR\ J0437$-$4715 by other groups, and present 
a possible interpretation, in Paper 2.

\section{Conclusions}

We find two scales of scattering for the nearby pulsar PSR\ J0437$-$4715.
The broader scale has scintillation bandwidth $\Delta\nu_{ISS}=16^{+8}_{-3}$\ MHz, and time scale $t_{ISS}=1000^{+350}_{-250}$\ sec.  
The narrower scale has 
$\Delta\nu_{ISS}=0.5\pm 0.1$\ MHz, and time scale $t_{ISS}=90\pm 20$ sec.  
Both yield scintillation velocities close to the proper-motion velocity found by van Straten (\cite{Van01}).
The pulse broadening time inferred from the broader scintillation bandwidth is 
less than that expected on the basis of observations of other nearby pulsars.

\begin{acknowledgements}
We thank the U.S. National Science Foundation for supporting this work.
\end{acknowledgements}

\clearpage

\begin{table}
\caption{
Measured Scintillation Parameters for PSR\ J0437$-$4715
}
\label{result}
\begin{tabular}{lcccc}
\hline\hline

      &\span\omit Wide-Bandwidth  &\span\omit Narrow-Bandwidth\\
Epoch &$\Delta\nu_{ISS}$           &$t_{ISS}$             &$\Delta\nu_{ISS}$ &$t_{ISS}$ \\
(MJD) &(MHz)              &(sec)               &(MHz)    &(sec)   \\

\hline     
50392        & $14.7 \phantom{{^{+5}_{-2}}^a} $& $\phantom{1}885 \phantom{{^{+350}_{-250}} ^{\ a}}  $&$0.6 \phantom{{\pm 20}^{\ b}} $&$ 90 \phantom{{\pm 20}^{\ b}} $\\
51276        & $16.5 \phantom{{^{+5}_{-2}}^a} $& $                  1630 \phantom{{^{+350}_{-250}} ^{\ a}} $&$0.4 \phantom{{\pm 20}^{\ b}}  $&$ 90 \phantom{{\pm 20}^{\ b}} $\\
Composite & $15.7 \phantom{{^{+5}_{-2}}^a} $& $                  1020 \phantom{{^{+350}_{-250}} ^{\ a}} $&$0.5 \phantom{{\pm 20}^{\ b}}  $&$ 90 \phantom{{\pm 20}^{\ b}} $\\
Adopted     & ${16^{+8}_{-3}}^{\ a}                  $&            ${1000^{+350}_{-250}} ^{\ a} $              &${0.5\pm 0.1}^{\ b}                   $&${90\pm 20}^{\ b} $\\
\hline
\end{tabular}
\par\noindent
$^{a}$ {Error estimated from Monte Carlo simulation.}
$^{b}$ {Error estimated from standard error from fits; see \S\ref{fine_scale_sect}.}
\label{results_table}
\end{table}


\begin{thebibliography}{}
\bibitem[2002]{cor02}Cordes, J.M., \& Lazio, T.J.W. 2002, astro-ph/0207156
\bibitem[1992]{Des92} Desai, K. M., Gwinn, C. R., Reynolds, J. R., King, E. A., Jauncey, D., Flanagan, C., Nicolson, G., Preston, R. A., \& Jones, D. L. 1992, \apjl, 393, L75
\bibitem[2000]{got00} Gothoskar, P., \& Gupta, Y. 2000, \apj, 531, 345 
\bibitem[1994]{Gup94} Gupta, Y., Rickett, B. J., \& Lyne, A. G. 1994, \mnras, 269, 1035 
\bibitem[1998]{Gwi98} Gwinn, C. R., Britton, M. C., Reynolds, J. E. J., Jauncey, D. L., King, E. A., McCulloch, P. M., Lovell, J. E., \& Preston, R. A. 1998, \apj, 505, 928 
\bibitem[2003]{hil03}Hill, A.S., Stinebring, D.R., Barnor, H.A., Berwick, D.E., \& Webber, A.B. 2003, ApJ, 599, 457
\bibitem[1998]{joh98}Johnston, S., Nicastro, L., \& Koribalski, B. 1998, MNRAS, 297, 108
\bibitem[1999]{lam99} Lambert, H. C., \& Rickett, B. J.  1999, \apj, 517, 299 
\bibitem[1995]{nic95}Nicastro, L., \& Johnston, S. 1995, \mnras, 273, 122
\bibitem[1992]{Phi92} Phillips, J. A., \& Clegg, A. W. 1992, \nat, 360, 137 
\bibitem[1997]{pyn97}Pynzar', A.V., and Shishov, V.I. 1997, Astronomy Reports, 41, 586
\bibitem[1982]{rob82}Roberts, J. A., \& Ables, J. G. 1982, \mnras, 201, 1119
\bibitem[2006]{smi06}Smirnova T.V., Gwinn, C.R., \& Shishov, V.I. 2005, \aa, submitted (Paper 2)
\bibitem[2001]{Sti01} Stinebring, D. R., McLaughlin, M. A., Cordes, J. M., Becker, K. M., Espinoza Goodman, J. E., Kramer, M. A., Sheckard, J. L., \& Smith, C. T. 2001, \apjl, 549, L97 
\bibitem[1993]{tay93} Taylor, J. H., Manchester, R. N., \& Lyne, A. G. 1993, \apjs, 88, 529 
\bibitem[1986]{Tho86} Thompson, A. R., Moran, J. M., \& Swenson, G. W., Jr. 1986, Interferometry and Synthesis in Radio Astronomy, (New York: John Wiley \& Sons) 
\bibitem[2001]{Van01} van Straten, W., Bailes, M., Britton, M., Kulkarni, S. R., Anderson, S. B., Manchester, R. N., \& Sarkissian, J. 2001, \nat, 412, 158 
\end{thebibliography}
\end{document}